\documentstyle[12pt]{article}

\input{psfig}

\def\picture #1 by #2 (#3){
  \vbox to #2{
    
\hrule width #1 height 0pt depth 0pt
    \vfill
    \special{picture #3} 
    }
  }

\def\scaledpicture #1 by #2 (#3 scaled #4){{
  \dimen0=#1 \dimen1=#2
  \divide\dimen0 by 1000 
\multiply\dimen0 by #4
  \divide\dimen1 by 1000 \multiply\dimen1 by #4
  \picture \dimen0 by \dimen1 
(#3 scaled #4)}
  }

\def\bphi1{\picture 3.50in by 5.00in (bphi1)}

\def\bphi2{\picture 3.50in by 5.00in (bphi2)}
\def\bphi3d{\picture 3.50in by 5.00in (bphi3d)}
\def\bkpi1
{\picture 3.50in by 5.00in (bkpi1)}
\def\bkpi2{\picture 3.50in by 5.00in (bkpi2)}

\def\etal{{\it et al.}}

\def\dI{{$\Delta I={1\over 2}\;$}}
\def\B{{{\cal{B}}_{\ell} (B) }}

\def\B{{{\cal B}_{\ell} (B) }}

\def\taur{{\tau (\Lambda_b)/\tau (B_d)}}

\def\ref#1{\cite{#1}}

\def\0{{\over}}
\newcommand{\bea}{\begin{eqnarray}}
\newcommand{\beq}{\begin{equation}}
\newcommand{\eea}{\end{eqnarray}}
\newcommand{\eeq}{\end{equation}}

\def\npb#1#2#3{    {Nucl. Phys. }{\bf B#1},
#3 (19#2)}
\def\plb#1#2#3{    {Phys. Lett. }{\bf #1B},
#3 (19#2)}
\def\prd#1#2#3{    {Phys. Rev. D}{\bf #1},
#3 (19#2)}

\def\prl#1#2#3{    {Phys. Rev. Lett. }{\bf
#1},
#3 (19#2)}

\def\zpc#1#2#3{    {Z. Phys. C}
{\bf #1},
#3 (19#2)}

\def\gtap{\ \raisebox{-.4ex}{\rlap{$\sim$}}
\raisebox{.4ex}{$>$}\ }

\begin{document}

\title{Hints for Enhanced $b \to s g $ from Charm and Kaon Counting} 

\author{Alexander L. Kagan \\[3mm] Department of
Physics \\ University of Cincinnati, Cincinnati, OH 45221\\[3mm]
Johan Rathsman\\[3mm] Stanford Linear Accelerator Center \\
Stanford University, Stanford,
California 94309}

\begin{titlepage} \maketitle

\begin{abstract} 

\thispagestyle{empty}

{Previously,  motivation for enhanced $b \to sg$ from new flavor physics has centered on discrepancies between theory and experiment.  Here two experimental hints are considered:  (1) updated measurements of the charm multiplicity and ${\cal B}(\overline{B} \to X_{c \bar c s})$ at the $\Upsilon (4S)$ imply ${\cal B}(B \to X_{no\;charm} ) \approx 12.4 \pm 5.6 \% $,  (2) the $\overline{B} \to K^- X$ and $\overline{B} \to K^+/K^- X $ branching fractions are in excess of conventional $\overline{B} \to X_c \to KX$ yields by about 
$16.9 \pm 5.6 \%$ and $18 \pm 5.3 \%$, respectively.  
JETSET 7.4 was used to estimate kaon yields from $s \bar s $ popping in $\overline{B} \to X_{c \bar u d}$ 
decays.  JETSET 7.4 Monte Carlos for ${\cal B}(\overline{B} \to X_{sg} ) \sim 15\%$
imply that the additional kaon production 
would lead to $1 \sigma$ agreement with observed charged and neutral kaon yields. 
The $K_s$ momentum spectrum would be consistent with recent CLEO bounds in the end point region.  
Search strategies for enhanced $b \to sg$ are discussed in light of large theoretical uncertainty 
in the standard model fast kaon background from $b \to s$ penguin operators. }

\end{abstract}

\end{titlepage}
\newpage

\section{Introduction}

Quark masses or Cabibbo-Kobayashi-Maskawa mixing angles arising from new interactions at the TeV scale are often 
directly correlated with large flavor changing chromomagnetic dipole moments since the chirality flip inherent in both sets of operators 
has a common origin \ref{kagan,holdom,wu}. 
Examples include radiatively induced quark masses via exchange of superpartners or vectorlike quarks \ref{kagan}, 
and dynamically generated quark masses in models with techniscalars \ref{kagan} or enhanced four fermion interactions \ref{holdom}.  
Another source for such correlations might be quark substructure \ref{peskin,luty}.  
In the case of enhanced $s \to dg$ a model-independent
analysis suggests that 20\% to 30\% of the $K \to \pi \pi $  \dI amplitude could be associated with generation of $\theta_c$ or $m_s$, with a 
corresponding scale $M$ for new physics below a TeV \ref{kagan}.  
Examples exist in which this does not lead to conflict with the neutral Kaon mass difference.  
This is important to keep in mind since large theoretical uncertainties currently make it 
impossible to know whether or not the standard model fully accounts for the observed \dI amplitude \ref{pich2}.

In the case of the $b \to sg$ dipole operators, a branching ratio of $\sim 10\%$ is typically associated with generation 
of $m_b$, $V_{cb}$ or $V_{ub}$ at scales near a TeV.  By way of comparison, in the standard model the $b \to sg $ BR is 
only $\sim 0.2 \%$ \ref{ciuchini}.
Some of the phenomenological consequences of enhanced $b \to sg$ for $B$ decays are decreases in the semileptonic branching ratio, $\B$, and
charm multiplicity , $n_c$ \ref{hou,bigi,kagan}, and increased kaon multiplicities.  
In fact, inclusive $B$ decay measurements currently hint at all three.  An updated comparison of $\B$ and $n_c$ with NLO predictions 
is the subject of a companion paper \ref{ncbsl} .  
We briefly summarize the current situation:  At $\mu \approx m_b$  higher order corrections to 
the $b \to c \bar u d $ decay width would have to be about 3.5 times than the NLO correction if perturbative QCD is 
to have a chance of simultaneously reproducing the low world averages for $\B$ and $n_c$ at the $\Upsilon(4S)$.  
Large negative corrections to the $b \to c \bar c s$ decay width would also be required but here perturbation theory appears to 
behave considerably worse \ref{lukeccs}.  

The discrepancy becomes considerably more serious if heavy quark effective theory [HQET] is to reproduce the low measured value 
of the lifetime ratio $\tau( \Lambda_b )/\tau(B_d) $ \ref{neubertsachrajda}. This necessarily requires negative spectator contributions to 
$\Gamma (B_d \to X_{c \bar u d} )$ at ${\cal O}(1/m_b^3)$, further increasing the theoretical prediction for $\B$. In this case higher order corrections 
to the $b \to c \bar u d $ decay width would probably have to be an order of magnitude larger than the NLO correction at $\mu \approx m_b $ \ref{ncbsl}.
The only standard model alternative is rather large deviations from local parton - hadron duality, amounting to a 10\% decrease in the $\Lambda_b$ decay width relative to the HQET prediction.
In a systematic attempt at classifying such deviations using a semi- quantitative one instanton gas approximation they have been found to be at the level of 1\% or smaller \ref{shifmannew}.  
On the other hand, $\B$, $n_c$ and $\taur$ are {\it{readily accomodated}} in HQET if the $b \to sg$ BR is $10 \%- 15 \%$.

In this paper we focus on charm and kaon counting in $B$ decays at the $\Upsilon(4S)$. In principle, the former can tell us the size of the charmless $b$ decay width,  while the latter can tell us how much of it is due to $b \to s $ transitions.  These are essentially experimental issues, 
relying much less on theoretical input than the analysis of $\B$ and $n_c$. 
We will see that the inclusive $\overline{B} \to K^- X $ and $\overline{B} \to K^+ /K^- X$ branching ratios \ref{ARGUSKaons,CLEOKaons} are 
approximately $3 \sigma$ to $3.5 \sigma$ in excess of the corresponding kaon yields from conventional 
$\overline{B} \to  X_c \to K X$ decays.
The latter are dominated 
by purely experimentally determined contributions, e.g., decays of intermediate $D$ or $D_s$ mesons. 
The most important unmeasured contribution to kaon production is
$s \bar s$ popping in $\overline{B} \to X_{c \bar u d}$ decays, which we estimate using 
the JETSET 7.4 \ref{jetset} string fragmentation model with 
both default and DELPHI tuning\footnote{We would like to thank Klaus Hamacher for providing us with the JETSET 7.4 DELPHI tunings, 
and Su Dong for incorporating them into our Monte Carlo.} \ref{hamdelphi}. Errors in the model are estimated by 
studying the dependence on a low mass cutoff for the initial
quark strings. Crude but generous estimates have also been included for kaon production from decays of intermediate charmed baryons and charmonium.

The excess in the charged kaon yield in $B$ decays, like the NLO analysis of $\B$ and $n_c$, suggests that the charmless $b \to s$ decay rate may be an order of magnitude larger than in the standard model \ref{kagantalks}.  In particular, we have studied kaon production from fragmentation in $\overline{B} \to X_{sg} $ decays using JETSET 7.4 with DELPHI tuning.  The result is that for ${\cal O}(15\% ) $ branching ratios the additional kaon yield leads to charged and neutral kaon multiplicities which are consistent with their measured values at the $1 \sigma $ level.  

It is important to note that enhanced $b \to sg$ is likely to be the only phenomenologically viable possibility for 
increasing kaon production in charmless decays.  For example, 10\% branching ratios for $b \to s q \bar q$ decays mediated by enhanced four quark operators 
\ref{hitoshi} typically violate the CLEO upper bound on high momentum $\phi$ production in $B$ decays \ref{CLEOphiep} by more than an order of magnitude \ref{kaganperez}.  Comparable branching ratios for $b \to s \nu \bar \nu $ from new physics \ref{holdom2} have been shown to be in gross violation of LEP constraints on missing energy in $b$ decays \ref{yuval}.   

Alternatively, it has been suggested that in $b \to c \bar c s$ decays, $c \bar c$ pairs in color octet states may annihilate often enough on a hadronic scale via a `long-range Penguin graph{'}
so that no new physics explanation would be necessary to account for the charm or kaon deficits  \ref{palmerstech,isinew}.    
However, this means that non-perturbative contributions to the charmless hadronic $b \to s$ branching ratio
would have to be of order 50 times 
larger than NLO corrections from penguin operator matrix elements
containing $c \bar c$ loops.  The latter causes shifts in the $b \to s q \bar q$ branching ratios of order
$2 \times 10^{-3} $ in magnitude  \ref{fleischer}.  
This possibility therefore seems remote, especially given that the relevant scale $\mu \sim 2 m_c $
is not all that small.  
It has also been suggested that ${\cal O}(1 /m_c^2 )$ corrections to the HQET $b$ 
dipole operator coefficient could lead to large corrections to the $b \to sg$ decay width 
\ref{voloshinnew}.   However, again a factor of 50 enhancement would be required which is unlikely
given that the corresponding correction to the $b \to s \gamma $ decay width was found 
to cause a change of only $ - 20 \%$.

The existence of a kaon excess was already noted in \ref{ARGUSKaons}.  However, at the time it was discounted as a signal for enhanced charmless $b$ decays since, among other things, flavor tagged measurements of inclusive $B \to D, \;D_s$ and $\Lambda_c $ decays 
were not yet available.  In fact, it was one of the motivations for the authors of Ref. \ref{isibuchalla} to suggest that 
${\cal B}(\overline{B} \to D \overline{D} \overline{K} X) $ is as large as $20\%$.  This decay has recently been measured \ref{CLEOBDDK}, and although indeed substantial has a rate which is about half as large as required in order to eliminate the $K^-$ excess. 
However, it almost doubles the measured $B \to X_{c \bar c s}$ BR without affecting the charm multiplicity.  As a result, an updated comparison of the two at the $\Upsilon (4S)$ via the identity 
\beq n_c = 1 + {\cal B} (B \to X_{c \bar c s}) - {\cal B} (B \to X_{{\rm{no \; charm}}}) \label{eq:ncbxg} \eeq 
gives a non-vanishing charmless branching ratio of about $12 \% $ at the $2 \sigma $ 
level \ref{kagantalks}.  This has also recently been discussed in
\footnote{By averaging this purely experimental method with a second method which 
derives from an essentially theoretical determination of ${\cal B}(B \to X_{c \bar u d})$, 
the authors of \ref{isinew} obtain a large $4 \sigma $ effect. The second method uses as input the NLO predictions 
for the ratio $\Gamma (b \to c \bar u d) / \Gamma(b \to c \ell \nu_{\ell})$.  We do not believe 
this procedure is justified given the remaining theoretical uncertainty in this ratio.  For example, 
if one includes vacuum polarization graph corrections to the semileptonic width resummed to all 
orders \ref{beneke}, the range of predictions for this ratio would be shifted 
upwards by ${\cal O} (10\%)$. Furthermore,
the purely experimental determination of ${\cal B}(B \to X_{c \bar u d})$, see Eq. 7, 
does not lead to a second `independent method{'}.}
\ref{isinew}.
Although the uncertainty is large,  
it is interesting that this result is consistent with the large $b \to sg$ branching fractions suggested by the $K^-$ excess and the 
low values of $\B$ and $n_c$.

The authors of Ref. \ref{milana} pointed out that it might be possible to see kaons from enhanced $b \to sg$ directly in the high 
momentum end point region, where there is little or no background from standard model $B \to X_c$ decays.  This is a difficult measurement because the large energy release in $b \to sg$ decays leads to high multiplicity final states from fragmentation \ref{swain}, or a soft kaon spectrum.
The CLEO collaboration has recently presented upper limits on inclusive 
$B \to K_s X $ branching ratios \ref{CLEOKsep} for $p_{K_s} > 2.1\;GeV$. 
The $\overline{B} \to X_{sg}$ Monte Carlo, with Fermi motion of the $b$ and spectator quarks included according to the model of Ref. \ref{altarelli}, 
implies that $K_s$ production in the end point region is consistent with these bounds for 10\% to 15\% branching ratios .  
Similar conclusions for $\phi $ production from fragmentation versus CLEO upper limits \ref{CLEOphiep} on high momentum $\phi$'s are presented in \ref{kaganperez,oldtalks}.

Our $\overline{B} \to X_{sg}$ Monte Carlo only takes into account string fragmentation at lowest order, i.e.,  parton showers have not been included.  In fact, there could be substantial destructive interference between those decays in which the gluon branches into a $q \bar q$ pair, i.e., 
$b \to s g^* \to s q \bar q$, and standard model decays via the penguin four quark operators $Q_3,..,Q_6$, depending on 
the phase of the $b_R \to s_L g$ chromomagnetic dipole operator.  As a result effective four quark operator contributions to high momentum kaon 
and $\phi$ production could be substantially smaller than in the standard model  \ref{kaganperez,oldtalks,ciuchini2}.  
Furthermore, standard model penguin operator contributions in the end point region are by themselves subject to large theoretical uncertainty.  
We will see that meaningful searches for fast kaons from enhanced $b \to sg$ must therefore include lower kaon momenta, e.g., $p_K \gtap 1.8\;GeV$ in 
the $\Upsilon(4S)$ rest frame.  As in searches restricted to higher momenta, the signal to background ratios would be about 1 to 1, but the latter 
would now be dominated by fast kaons from cascade $b \to c \to s$ decays.  This contribution should, in principle, be possible to pin down 
experimentally to high precision by combining measurements of inclusive $D,D_s$ and $K$ momentum spectra from $B \to D/D_s $ and $D/D_s \to K $ 
decays, respectively \ref{milana}.  We will make use of kaon spectra from SLD's implementation of the CLEO B decay Monte Carlo.\footnote{We would like to thank Su Dong for providing us with the results of this simulation.}

\vspace*{0.3cm} 

This paper is organized as follows: In the next section we
review the experimental inputs relevant to inclusive charm and kaon counting in $B$ decays.  In Section 3 we determine the kaon yields from $\overline{B} \to X_c$ decays and compare with the total kaon yields.  Section 4 discusses kaon production and the $K_s$ momentum spectrum from fragmentation of enhanced $b \to sg$.  We conclude with a brief discussion of our results in Section 5

\section{Experimental input from inclusive $B$ decays}

{\noindent \it{Counting charm I:  the charm multiplicity}}

\vspace*{0.4cm}

The inclusive $B$ to charmed hadron branching ratios 
used to obtain the $B$ decay charm multiplicity at the $\Upsilon(4S)$ are given in Table 1.
The first four entries are averages of the ARGUS, CLEO 1.5 and CLEO II measurements, 
from Ref. \ref{honscheid}.\footnote{We've taken $ .6 \pm .3 $ for the sum of all unknown charmonia yields.  
Currently, $ {\cal B}(B \to \eta_c X) < .9 \% $ ($90\%$  C.L.) [37].} 
However, the $D^0/ \overline{D^0} $ entry has been rescaled to the new 
$ D^0 \to  K^- \pi^+ $ world average\footnote{This includes the new ALEPH measurement presented at the summer conferences, ${\cal B} (D^0 \to  K^- \pi^+ ) = 3.90 \pm .15 \%$ \ref{ALEPHDKPi}.  ${\cal B} (B \to  D^0/\overline{D^0} X)$ is inversely proportinal to ${\cal B} (D^0 \to  K^- \pi^+ )$, which was taken to be 
$3.76 \pm .15 \%$ in \ref{honscheid}.  Similarly,   ${\cal B}(B \to D^+/ \overline{D^-} X$ is inversely proportional to 
 ${\cal B} (D^+ \to K^- \pi^+ \pi^- ) $ which was taken to be $8.9 \pm .7 \%$ in \ref{honscheid}.}
branching ratio \ref{richman}, $3.88 \pm .10 \%$, and  the $D^+/ \overline{D^-} $ entry has been rescaled to the
 $D^+ \to K^- \pi^+ \pi^-  $ PDG 96 branching ratio \ref{PDG96} , $9.1 \pm .6 \% $.

The CLEO $\Lambda_c$ yield in Table 1 \ref{yamamoto} is obtained from a direct measurement of ${\cal B}(B \to p K^- \pi^+ X)$ \ref{zoeller}. 
The largest
uncertainty is due to ${\cal B}(\Lambda_c \to p K^- \pi^+ )$, which has been set equal to the PDG 96 average of $4.4 \pm .6 \%$.  However, 
it has been observed that the latter is largely based on a flawed model of baryon production in $B$ decays \ref{isilong,honscheid}, namely dominance 
of the external $W$ spectator diagrams in baryon production.  This model can not be correct given the absence of a signal 
for $\overline B \to \Lambda_c \overline{N} X \ell \nu $ \ref{jain,CLEOLambdatag}.   An alternative method used by CLEO \ref{CLEObergfeld,honscheid} 
combines a measurement of the relative semileptonic rate, $\Gamma(\Lambda_c^+ 
\to p K^- \pi^+ ) / \Gamma(\Lambda_c \to \Lambda \ell^+ \nu_{\ell} )$,
with an additional assumption about the fraction of $\Lambda_c \to \Lambda \ell^+ \nu $ decays versus all semileptonic $\Lambda_c$ decays.  
The resulting $\Lambda_c \to p K^- \pi^+ $ branching ratio is higher than the PDG 96 average, which would lower the $\Lambda_c $ 
yield.\footnote{Using this method gives
${\cal B}(\Lambda_c \to p K^- \pi^+ ) =  5.9 \pm 1.5 \%   $, leading to ${\cal B }(B \to \Lambda_c /\overline{\Lambda_c} X) = 3.1 \pm 1.0 \% $.}    

The $\Xi_c$ yield has large uncertainties and the central value appears large compared to the $\Lambda_c $ yield.  
In Ref. \ref{isilong} the $\Xi_c$ yield has been correlated with the more accurately measured $\Lambda_c $ yield.  Allowing for a
probability for $s \bar s $ popping from the vacuum of $15 \pm 5 \%$ leads to an estimate for the ratio of the $\Xi_c $ yield
to the $\Lambda_c $ yield of $ .41 \pm  .12$.  Combining this with the $\Lambda_c $ entry in Table 1 
leads to a significantly smaller $B \to \Xi_c X$ branching ratio of
$1.7 \pm .6 \% $.   Note that in this case the sum of the $\Lambda_c $ and $\Xi_c$ yields is in better agreement 
with previous measurements 
of the total charmed baryon multiplicity \ref{CLEObarctot},  $6.4 \pm 1.1 \%$, and total baryon mulitiplicity 
\ref{ARGUSbartot}, $6.8 \pm .6 \%$.

\begin{table}\begin{center}
\caption{Inclusive $ B \to\;charmed\;hadron $, and $ B \to K $ BR's [\%]. $(c \bar c)$ is any $c \bar c$  meson.   
The second $ \Xi_c $ entry is correlated with the $ \Lambda_c $ yield, as in Ref. [43].} 
\vspace*{0.5cm} 
\begin{tabular}{| c | c |}\hline
Process  & Branching Ratios  \\
\hline
$\overline{B} \to D^0 / \overline{D^0} X $ & $62.8 \pm 2.7  $    \ref{honscheid}\\
$\overline{B} \to D^+ /  D^- X $ &  $23.7 \pm 2 $ \ref{honscheid}\\
$\overline{B} \to D_s^+  / D_s^- X $ &  $10.1 \pm 2.6 $    \ref{honscheid}   \\
$\overline{B} \to (c \bar c) X_s $ .
   & $ 2.6 \pm .3  $ \ref{honscheid}\\
$\overline{B} \to \Lambda_c^+  / \Lambda_c ^- X $ & $ 4.1 \pm .6 $ \ref{yamamoto}\\
$\overline{B} \to \Xi_c^+ /\Xi_c^0  X $ & $ 3.9 \pm 1.5 $ \ref{yamamoto}\\
(Correlated $\Xi_c$ yield)& $ 1.7 \pm .6  $ \\
\hline
$\overline{B} \to K^- /K^+ X $ & $78.9 \pm 2.5$~\ref{PDG96}  \\
$\overline{B} \to K^0 /\overline{K^0} X $ &  $64  \pm 4$~\ref{PDG96} \\ 
\hline 
\end{tabular}
\end{center}
\end{table}
The world-average $B$ decay charm multiplicity at the $\Upsilon(4S)$ is given by the sum of the six
charmed hadron yields in Table 1, with the $(c \bar c)$ yield
counted twice.  The result is
\beq n_c (B)= 109.8 \pm 4.6 \% .\label{eq:nchigh} \eeq
Using the correlated $\Xi_c $ yield in parenthesis leads to 
\beq n_c (B) = 107.6 \pm 4.4 \%. \label{eq:nclow} \eeq  
Note that uncertainty in the absolute $D$ and $D_s$ branching scales contributes about $\pm 3.4 \%$ to the error in $n_c$.
Finally, the CLEO II charm multiplicity alone is 
$113.4 \pm 4.6 \% $ \ref{yamamoto}, with a correspondingly lower value if the correlated
$\Xi_c $ yield is used.  
For the admixture of beauty hadrons at the $Z$ the reported charm multiplicities are
$123 \pm  7.5 \% $ at ALEPH \ref{ALEPHnc} and $110 \pm 8.8\%$ at OPAL \ref{OPALnc}, where the latter does not include the $\Xi_c$ contribution.

\vspace*{0.4cm}

{\noindent \it{Counting charm II: the $B \to X_{c \bar c s} $ and $B \to X_{c \bar u d}$ branching ratios.}}

\vspace*{0.4cm} 

The flavor  - lepton charge correlation measurements  
for inclusive $B$ to charmed hadron decays are summarized in Table 2.\footnote{The relative $K^+$ and $K^-$ yields follow from
the ARGUS and CLEO flavor - lepton charge correlation measurements \ref{ARGUSKaons,CLEOKaons,IsiKtag}.  The
effect of $B - \overline{B}$ mixing on the CLEO measurement is accounted for in \ref{IsiKtag}.  
The absolute $K^+$ and $K^-$ BR's are obtained from the total charged kaon yield in Table 1.} The flavor tagged BR{'}s are obtained by combining the CLEO collaboration{'}s measurements of the relative flavor yields in the first column
with the averages for the total $D_s$, $D$ and $\Lambda_c$ BR{'}s from Table 1.
They can be used to obtain
\bea {\cal B} (B \to X_{c \bar c  s}) & = & {\cal B} (\overline{B} \to D_s^- X)
+ {\cal B} (\overline{B} \to \overline{D} X) + {\cal B} (\overline{B} \to \overline{\Lambda_c}^- X) \nonumber \\
&  & + {\cal B} (\overline{B} \to (c \bar c) X) = 20.0 \pm 3.5 \%. \label{eq:BtoXcc} \eea
Note that the $\overline{\Lambda_c}^- $ yield is identified
with the $\Xi_c$ yield from $B  \to \Xi_c \overline{\Lambda_c}^- X $ decays.

The $\overline{B} \to X_{c \bar u d} \to DX / D_s X $ branching ratio\footnote{$X_{ c \bar u d}$ is understood to include the Cabbibo suppressed final states $X_{ c \bar u s}$.}  will be required in order to estimate contributions to kaon production from $s \bar s $ popping.  It is obtained from the total $D$, $D_s$ and charmed baryon yields in Table 1 by subtracting the contributions from the semileptonic and $b \to c \bar c s $ decays. 
The sum of the $D$ and $D_s$ yields originating from semileptonic decays can, to good approximation, be identified with the sum of the PDG 96 average branching ratios for $B \to e \nu_e X,\;\mu \nu_{\mu} X$, and $\bar b \to \tau^+ \nu_{\tau} X $, giving ${\cal B}(B\to X_{c \ell \nu_{\ell} } \to DX,D_s X )\approx 23.4 \pm .8 \% $.  The charmed baryon yield from semileptonic decays is neglected. 
The sum of all $D$ and $D_s $ meson yields from $\overline{B} \to X_{c \bar u d}$ decays is
\bea {\cal B}(\overline{B} \to X_{c \bar u d} \to DX/D_s X)& = &
{\cal B}(\overline{B} \to D^0 / \overline{D^0} X) + {\cal B}(\overline{B} \to D^+ /  D^- X) \nonumber \\
- {\cal B}(\overline{B}\to X_{c \ell \nu_{\ell}} \to D/D_s X )
& - & 2 {\cal B}(\overline{B} \to \overline{D} X ) - 
{\cal B}(\overline{B} \to D_s^- X )  \nonumber \\  
+ {\cal B}(\overline{B} \to D_s^+ X )& = & 39.6 \pm 6.7 , \label{eq:Dcud} \eea
The $D$ yield alone, obtained by eliminating the last term above, is 
$37.9 \pm 6.6 \%$.

Similarly, the sum of the $\Lambda_c $ and $\Xi_c$ yields from $b \to c \bar u d$ decays is given by
\bea {\cal B}(B \to X_{c \bar u d} \to \Lambda_c X, \Xi_c X) & =
& {\cal B}(\overline{B} \to \Lambda_c/\overline{\Lambda_c} X) +
{\cal B}(\overline{B} \to \Xi_c^+ /\Xi_c^0  X  ) \nonumber \\ 
- 2 {\cal B}(\overline{B}
\to \overline{\Lambda_c}^- X) & = & 
6.6 \pm 1.8 \% ,\;4.4 \pm 1.2 \% . \eea
Summing Eqs. 5 and 6 gives the total inclusive branching ratio
\beq {\cal B}(B \to X_{c \bar u d} ) =   46.2 \pm 7.0\%,\;44.0 \pm 6.8\% . \label{eq:Xcud} \eeq
The second entry in Eqs. 6 and 7 is obtained using the correlated $\Xi_c$ yield in Table 1.

\begin{table}\begin{center}
\caption{
Inclusive flavor tagged $B$ decay branching ratios [\%]. 
$D$ denotes $D^0$ or $D^+$, and similarly for $\overline{D}$. }

\vspace*{0.5cm} 

\begin{tabular} {| c | c | c | c |} \hline
$T$  & ${{\cal B}(\overline{B} \to \overline{T}X) \over  {\cal B}(\overline{B} \to TX) } $ &    
${\cal B}(\overline{B} \to \overline{T}X)$  &   ${\cal B}(\overline{B} \to TX) $ \\
\hline
$D$ &  $.107 \pm  .034$
 \ref{CLEOBDDK}   & $8.4 \pm 2.6 $      & $78.1 \pm 3.8 $ \\
$D_s^- $ & $.21 \pm .10 $ \ref{CLEODstag}  & $1.74 \pm 1.0$ & $8.36 \pm 2.3$ \\
$\Lambda^+_c $ &  $.20 \pm .13 $ \ref{CLEOLambdatag}      &    $.68 \pm .45 $ & 
$3.42 \pm .62 $   \\
\hline
$K^- $ & $ .18 \pm .06$  &  $12.0 \pm 3.7 $ & $ 66.9 \pm 3.8 $ \\
\hline
\end{tabular}
\end{center}
\end{table}

\vspace*{0.3cm}

{\noindent \it{A bound on $b \to sg$ }}

\vspace*{0.4cm}

The charm multiplicity and ${\cal B}(\overline{B} \to X_{c \bar c s})$ can be used to bound ${\cal B} (\overline{B} \to X_{sg})$ via
Eq. 1.
The $\Upsilon(4S)$ measurements given in Eqs. 2 - 4
lead to                
\beq {\cal B} (B \to X_{\rm{no\;charm}}) = 10.2 \pm 5.8\% , \;12.4 \pm 5.6 \%    \label{eq:bxg1} \eeq
where the second entry again corresponds to the correlated $\Xi_c $ yield in Table 1.   Bounds on ${\cal B}(B \to X_{sg})$ follow by subtracting $\approx 1 \%$ to account for $b \to u$ decays.\footnote{Standard model $b \to s q \bar q$ penguin operator contributions should not be subtracted since, as noted in the Introduction, enhanced $b \to sg$ can destructively interference with these decays via gluon splitting.}
This method should ultimately provide one of the most accurate determinations of the charmless branching ratio
at the $B$ factories if uncertainties in the absolute $D$ and $D_s$ branching scales are substantially reduced. 
In the meantime, the resulting bounds are consistent with either enhanced $b \to sg$, or no $b \to sg$.

\section{Counting kaons in $B$ decays}

In this section we would like to check for enhanced charmless $b \to s$ transitions by comparing the inclusive $\overline{B} \to K^- X$, $\overline{B} \to K^+ X$ and $\overline{B} \to K^0 /\overline{K^0} X$ BR{'}s given in Tables 1 and 2 with the corresponding kaon yields from $\overline{B} \to X_c$ decays.  Contributions to the latter which are essentially determined experimentally are summarized in Table 3.  Lets begin with contributions from decays of intermediate $D$ mesons, which are obtained by combining inclusive $\overline{B} \to D X$ BR{'}s with 
the corresponding inclusive PDG 96 $D \to K X$ BR{'}s.   
To obtain the $K^-$ and $K^+$ yields separately requires knowledge of the individual charged and neutral $ D$ and $\overline D$ multiplicities in $\overline{B}$ decays, which in turn requires knowledge of the $\overline{D^0}$ and $D^-$ components of the `wrong flavor{'} $\overline{B} \to \overline{D} X$ BR in Table 2. 
The following partial results have recently been reported by the ALEPH collaboration \ref{ALEPHDDbar}
\bea {\cal B}(B^0,B^{\pm} \to D^0 \overline{D^0} X ) & = & 7.6 \pm 2.55 ^{+0}_{-0.6} \% \nonumber\\
{\cal B}(B^0,B^{\pm} \to D^{\pm} \overline{D^0} X ) & = & 5.2 \pm 2.25 ^{+0}_{-0.4} \% ,
\eea
indicating that most $\overline{D}${'}s produced in $\overline{B}$ decays are neutral.  This is to be expected since $D^*${'}s decay preferentially to neutral $D${'}s.  Fortunately, the charged kaon yields are not very sensitive to the ratio of charged to neutral $\overline{D}$ multiplicities (the total neutral kaon yield is independent of it), since most $D$ mesons produced in $\overline{B}$ decays are $D^0${'}s or $D^+${'}s.  In particular, the
$K^-$ and $K^+$ yields
vary from $36.2 \pm 3.0 \%$ and $ 6.4 \pm 1.0 \%$ to $34.9 \pm 3.1 \%$ and
$7.7 \pm 1.5 \%$, respectively, as this ratio is varied from 1 to 0. 
The corresponding charged kaon entries in Table 3
are for the intermediate case where $\overline{D^0}$ and
$D^-$ are assumed to account for 75\% and 25\% of the $\overline{B} \to \overline{D} X$ BR, respectively. 

Kaon yields originating from the decay of $ D_s$ intermediaries are straightforward to obtain by combining the $D_s$ entries in Table 2 with the PDG 96 $D_s \to
KX$ BR{'}s.  Two other sources for kaons whose contributions follow directly from Tables 1 and 2 are $\overline{B} \to D \overline{D} (X_s \to K^- X,\overline{K^0} X)$
and $\overline{B} \to (c \bar{c}) (X_s \to K^- X,\overline{K^0} X)$, where $(c \bar c)$ denotes all charmonium bound states.\footnote{Hadronization of the $s$ quark
into strange baryons is kinematically forbidden in the first decay and can be neglected in the second decay due to phase space suppression.}  In each decay the probability that the $s$ quark hadronizes into
$K^-$ or $\overline{K^0}$ should be near 50\%, given an equal admixture of $\overline{B^0}$ and $B^-$, apart from small isospin breaking effects.  Finally, another source of kaons in $\overline B$ decays on which we have a good handle is
hadronization of the initial $s $ quark in Cabibbo suppressed $\overline{B} \to X_{c \bar u s} \to D K X$ decays, which is expected to contribute about 1\% to the $K^-$
and $\overline{K^0}$ multiplicities.  From Eq. 5
\beq {\cal B}(\overline{B} \to X_{c \bar{u} s} \to D K X) \sim \theta_c^2 {\cal B}(\overline{B} \to X_{c \bar{u} d} \to D X) = 1.8 \pm .3 \% .\eeq 
Hadronization of the $s$ quark
should lead to approximately equal numbers of $K^-$ and $\overline{K^0}$ since the large energies of the primary $s$ and $\bar u$ quarks 
lead to high probabilities for $q \bar q$ popping along initial $[s \bar u]$ strings.

The above contributions to kaon production have been summed in Table 3.  Comparison with the inclusive kaon yields in Tables 1 and 2 
shows that at this stage there exist sizable $4.4 \sigma $ and $5.6 \sigma$ excesses in $K^-$ and $K^+/ \overline{K^-}$ production, 
respectively, which must be accounted for:
\bea {\cal B}(\overline{B} \to K^- X) - {\cal B}(\overline{B} \to X_c \to K^- X)_{\rm{measured}}
 &=  &23.1 \pm 5.3 \nonumber\\
\!\!\!{\cal B}(\overline{B}\! \to \! K^+ /K^- X)\! - \!
{\cal B}(\overline{B} \to X_c \! \to \! K^+ /K^-  X)_{\rm{measured}} &\! = \! & 26.7 \pm 4.8
 \label{eq:Kexcessstage1} \eea

\begin{table}\begin{center}
\caption{
Known contributions to inclusive kaon multiplicities [\%] from $\overline{B} \to X_c$ decays.}
\vspace*{0.5cm} 
\begin{tabular} {| c | c | c | c |} \hline
Decay Mode  & $K^- $ &    
$ K^+ $ &   $K^0 / \overline{K^0}$ \\
\hline
$\overline{B} \to D,\overline{D} \to K$  & $ 35.5 \pm 3.0     $      & $7.0 \pm 1.2$ &  $40.3 \pm 3.9 $     \\
$\overline{B} \to D^+_s , D^-_s \to K$ & $ 1.9^{+ 1.6}_{-1.3}  $ & $1.4 ^{+1.3}_{-1.1}$ & $3.9 \pm 3 $\\
$ \overline{B} \to D \overline{D} (X_s \to K)$ & $4.2 \pm 1.3$ & - & $4.2 \pm 1.3$ \\
$ \overline{B} \to (c \bar c) (X_s \to K) $ & $ 1.3 \pm .2 $ & - & $1.3 \pm .2 $ \\
$\overline{B} \to X_{c \bar{u} s} \to D K X $ & $.9 \pm .2 $ & - & $.9 \pm .2 $ \\
 \hline
totals & $43.8 \pm 3.7 $ & $8.4 \pm 1.8 $ & $50.6  \pm 5.1 $ \\
\hline
\end{tabular}
\end{center}
\end{table}

It remains to estimate those contributions to kaon production for which there is limited experimental information, 
namely $s \bar s$ popping in $b \to c \bar{u} d$ decays\footnote{We will not bother to consider $s \bar s$ popping in $b \to c \bar u s$ decays separately as this would have negligible consequences for our purposes.}, 
and decays of intermediate charmed baryons and charmonium.  For a summary see Table 5.  
In the case of $s \bar s $ popping additional kaons are most likely to come from final states of the form $D K \overline{K} X$ and, 
to a lesser extent, $D_s \overline{K} X$.  Note that $s \bar s$ popping in $\overline{B} \to \Lambda_c $ 
decays can be safely neglected for our purposes due to the small overall $\overline{B} \to \Lambda_c X $ BR.  
In $b \to c\bar c s$ decays it can be neglected because of phase space suppression, whereas in semileptonic decays 
it can be neglected because it constitutes a small fraction\footnote{The sum of the $b \to c e \nu_e$ and $b \to c \mu \nu_{\mu}$ BR{'}s is about half as large as the $b \to c \bar u d$ BR, and the invariant mass of the $c \bar q$ string  in semileptonic decays is typically too small for $s \bar s$ popping because of the low relative momenta of the charm and spectator quarks.} of an already small overall $\overline{B} \to D^+_s X$ BR, see Table 2.

A rough estimate of the kaon yield from $s \bar s$ popping can be obtained by assuming that the latter occurs with a probability $P_s$ of $15 \pm 5 \% $ in $\overline{B} \to X_{c \bar u d} \to DX/D_s X$ decays, i.e.,
\beq P_s \equiv {{\cal B} (\overline{B} \to X_{c \bar u d + s \bar s} \to D K \overline{K} X/D_s \overline{K} X) \over  
{\cal B} (\overline{B} \to X_{c \bar u d } \to DX /D_s X )} = 15 \pm 5 \%, 
\label{eq:ssrough} \eeq 
which is typical of $s \bar s$ popping probabilities in the literature for various $B$ decays.  In turn, Eq. 5 gives
${\cal B} (\overline{B} \to X_{c \bar u d + s \bar s} \to D K \overline{K} X/D^+_s \overline{K} X) \approx  5.9 \pm 2.2 \% $.
Estimates for the individual kaon yields follow by assuming that the $s$ and $\bar s$ quarks hadronize into charged and neutral kaons with equal probability.  This is approximately true if $s \bar s$ popping is accompanied by popping of several light quark pairs, which is certainly the case along the more energetic initial $[\bar u d]$ string chiefly responsible for $s \bar s$ popping.  The individual kaon yields implied by Eq. 12 would be 
\bea {\cal B}(\overline{B} \to D/D^+_s \, K^- X) \approx 3.0 \pm 1.1 \% &,&{\cal B}(\overline{B} \to D K^0 X, D/D^+_s \,\overline{K^0} X) \approx 5.0 \pm 1.6 \% \nonumber\\ 
 {\cal B}(\overline{B} \to D K^+ X) & \approx & 2.1 \pm 1.2 \%, \label{eq:sscrude} \eea
where ${\cal B}(\overline{B} \to D_s^+ X )$ has been subtracted in obtaining the $K^0 $ and $K^+$ estimates.

Next we estimate the kaon yield from $s \bar s$
popping using JETSET 7.4 string fragmentation at lowest order.  
Simple color counting arguments with no additional dynamical factors taken into account, e.g.,  a candidate string¹s invariant mass,
imply that of the two possible initial string configurations, $[c \bar q]$ + $[\bar u d]$ and $[c \bar u]$ + $[d \bar q]$ ($\bar q$ is the spectator quark), the former is about five times more likely.\footnote{Color counting gives a 5 to 1 ratio for the dominant four quark operator $Q_2$.  Including the contribution of $Q_1$ gives a leading order ratio which varies from $\approx 5.5$ at $\mu = m_b$ to $\approx 4.5$ at $\mu = m_b /2$.}  Taking a 5 to 1 ratio, the default JETSET 7.4 settings give 
an $s \bar s$ popping probability $P_s$ (see Eq. 12) of $11 \% $.  The DELPHI tuning gives a
larger probability of 14 \%, presumably the result of improved agreement with observed kaon production at the $Z$.
Using Eq. 5 to set the absolute branching scale 
leads to the kaon and $D_s^+$ yields listed in Table 4, with errors corresponding to the experimental uncertainty in Eq. 5.

Admittedly, initial string invariant masses in $\overline{B} \to X_{c \bar u d}$ decays are on the low side for a jet-like interpretation.  
To get an idea of the corresponding uncertainty in the $s \bar s$ popping probability, or of how much larger the resulting kaon yields could be, we have repeated the analysis with a lower cutoff of 2 GeV imposed on the invariant masses of the initial $[\bar u d]$ or $[\bar u c]$ strings along which most of the $s \bar s$ popping 
is expected to occur  ($\approx 70\% $ of the decays survive this cut).\footnote{The same cut is applied to any candidate string¹s invariant mass 
in the JETSET $e^+ e^- $ continuum package, and to $gg$ invariant masses in the JETSET $\Upsilon$ decay package \ref{jetset}.}
These results are also included in Table 4; the corresponding values of $P_s$ are $13\% $ and 
17\% for the default and DELPHI JETSET tunings, respectively.  
Note that the range of kaon yields in Table 4 is consistent with the rough estimates in Eqs. 12 and 13.  
When evaluating the charged kaon excess below we will equate the central values of the kaon yields from $s \bar s$ popping with the central values of the 
DELPHI tuned entries obtained without a string mass cutoff. Error bars are determined by the maximum kaon yields attainable with 2 GeV cutoff. 
These are included in Table 5.
As a rough check we note that the range of $ D_s^+ $ yields in Table 4, although on the low side, is consistent with the poorly measured $\overline{B} \to D_s^+ X$ BR in Table 2.  It is also worth noting that JETSET estimates for kaon production in $\Upsilon(1S)$ decays are consistent with the measured multiplicity for charged kaons and within 20\% for neutral kaons.\footnote{The quarkonia decay subroutine LUONIA gives charged and neutral kaon multiplicities of .9 and .84, respectively, whereas the ARGUS measurements \ref{arguskfrag} are $.908 \pm  .026$ ($K^+ /K^- $)
and $1.033 \pm .05$ ($K^0 /\overline{K^0}$).}

\begin{table}\begin{center}
\caption{Lowest order JETSET 7.4 estimates for kaon and $D_s^+ $ multiplicities [\%] from $s \bar s$ popping in decays of the form $\overline{B} \to X_{c \bar u d + s \bar s} \to DK \overline{K}X / D_s^+ \overline{K} X$.  In the second and fourth rows a 2 GeV lower cutoff has been imposed on the invariant masses of the $[\bar u d]$ or $[c \bar u]$ strings.}
\vspace*{0.5cm} 
\begin{tabular} {| c | c | c | c | c |} \hline
JETSET 7.4 settings  & $K^- $ &    
$ K^+ $ &   $K^0 / \overline{K^0}$ & $D_s^+  $\\
\hline
default  & $ 2.6 \pm .4     $      & $1.4 \pm .2 $ &  $3.8 \pm .6 $ 
& $.6  \pm .1 $    \\
default, 2 GeV cut  & $ 3.2 \pm .5 $ & $1.8 \pm .3 $ & $ 4.7 \pm .8$ 
&  $.9 \pm  .2 $ \\
DELPHI tunings & $3.7 \pm .6 $ & $1.6 \pm .3$  & $4.9 \pm .8 $ &
$ .5 \pm .1 $\\
DELPHI, 2 GeV cut  & $4.6 \pm .8 $ & $2.1 \pm .4 $ & $6 \pm 1$  & $.8 \pm .1 $ \\
 \hline
\end{tabular}
\end{center}
\end{table}

Finally, we crudely estimate kaon production from intermediate $\Lambda_c$, $\Xi_c$ and charmonium decays.   
The $\Lambda_c^+ \to  \Lambda X$ and $\Lambda_c^+ \to  \Xi^{\pm} X$ BR{'}s are $35 \pm 11 \%$ and $10 \pm 5\%$, respectively.
The remaining $\Lambda_c ^+ $ decay modes give an upper bound on the
$\Lambda^+_c \to \overline{K} X$ BR of $55 \pm 12 \%$,
which we assume is saturated.
It is far from clear what percentage of kaons will be charged or neutral, as this depends on the importance of light quark popping in kaon production, and the probability of $\overline{K^{*0}}$ versus $\overline{K^0}$ production from the primary $s$ and $\bar u$ quarks.   
For lack of anything better we take 
the $\overline{K}${'}s to be $50 \pm 25 \%$ neutral and $50 \pm 25\%$ charged, with large error bars to reflect our ignorance.  
In turn, the flavor tagged $\Lambda_c $ yields in Table 2 lead to the $\overline{B} \to \Lambda_c \to K$ estimates in Table 5.
Even less is known about inclusive $\Xi_c$ decays.  
It is reasonable that the relative probability of the decay chain $c \to s \to \overline{K}$ is roughly 
the same as in $\Lambda_c$ decays, based on approximate flavor $SU(3)$;  We choose $50  \pm 20\% $.
Lets also assume that the spectator $s$ quark
hadronizes into a kaon $50 \pm 25 \% $ of the time, corresponding to a total kaon multiplicity of $\approx 100\%$.  
Again we take the kaons to be $50 \pm 25\%$ charged and $50 \pm 25\%$ neutral.  The $K^+$ and $K^0$ yields are negligible because $\overline{\Xi_c}$ production is highly suppressed in $\overline{B}$ decays.

For intermediate charmonium decays a crude estimate of the kaon yields is obtained by assuming a 
$50 \pm 25\%$ $s\bar s$ popping probability in the dominant $(c \bar c) \to ggg$ hadronic decay modes.  
It should be smaller than the $s \bar s$ popping probability in $\Upsilon(1S) $ decays which is about 100\% \ref{arguskfrag} because of 
the lower energy release.\footnote{For example, the JETSET 7.4 LUONIA subroutine applied to $J/\Psi $ decays, 
only to be regarded as an extremely rough estimate because of the low string energies involved, gives a $35\%$ $s \bar s$ popping 
probability and $\approx .18$ for the four individual kaon muliplicities.}   The crude charmonium estimates in Table 5 follow from 
the inclusive charmonium yield in Table 1 by assuming that the $s \bar s$ pairs always fragment into kaons.  
The kaons should be $\approx 50 \%$ neutral and $50 \%$ charged as in $\Upsilon(1S)$ decays \ref{arguskfrag}. 

\begin{table}\begin{center}
\caption{
Estimates of inclusive kaon yields [\%] from $\overline{B} \to X_c$ decays which have not been measured, 
as described in the text.  $s \bar s $ popping contributions are from the JETSET 7.4 Monte Carlo analysis, see text for details.  Estimates obtained using the correlated
$\Xi_c$ yield in Table1 are also included.}
\vspace*{0.2cm} 
\begin{tabular} {| c | c | c | c |} \hline
Decay Mode  & $K^-  $&    
$ K^+ $ &   $K^0/ \overline{K^0}$ \\
\hline
$\overline{B} \to X_{c \bar u d + s \bar s} \to DK\overline{K} X, D^+_s \overline{K} X $  & $ 3.7  \pm  1.7  $     &  $1.6 \pm .8 $ &  $ 4.9 \pm 2.1 $     \\
$\overline{B} \to \Lambda_c  \to  K$ & $ .9 \pm .5 $ & $.2 \pm .1$ & $1.1 \pm .6 $\\
$ \overline{B} \to \Xi_c^+ /\Xi_c^0 \to K $ & $  2.0 \pm 1.4$ & -  & $ 2.0 \pm 1.4$ \\
(Correlated $\Xi_c$ yield) & $ .9 \pm .6$ & - & $ .9 \pm .6 $ \\
 $\overline{B} \to (c \bar c) \to K$ & $  .7 \pm .3$ &$.7 \pm .3$ & $  1.3 \pm .3$ \\
\hline
Totals & $ 7.3 \pm 2.3 $ & $ 2.5 \pm .9 $ & $9.3 \pm 2.6 $ \\
(Correlated $\Xi_c$ yield)& $6.2 \pm 1.9 $ &$2.5 \pm .9 $ & $8.2 \pm 2.3$ \\
\hline
\end{tabular}
\end{center}
\end{table}

Bounds on kaon production from charmless intermediate states are obtained by subtracting the totals in Tables 3 and 5 from the inclusive $\overline{B} \to K X$ BR{'}s, giving [\%]
\bea {\cal B}(\overline{B} \to K^- X) - {\cal B}(\overline{B} \to X_c \to K^- X)
& =  &15.8 \pm 5.8,\;16.9 \pm 5.6 \nonumber\\
{\cal B}(\overline{B} \to K^+ X) - {\cal B}(\overline{B} \to X_c \to K^+ X)
& =  & 1.1 \pm 4.2,\; 1.1 \pm 4.2 \nonumber\\
{\cal B}(\overline{B} \to K^+ /K^- X) - 
{\cal B}(\overline{B} \to X_c \to K^+ /K^- X) & = & 16.9 \pm 5.4,\;18 \pm 5.3 \nonumber \\
\!\!\!{\cal B}(\overline{B}\! \to\! K^0 /\overline{K^0} X)\! - \!
{\cal B}(\overline{B}\! \to\! X_c \!\to \!K^0 /\overline{K^0} X) &\!\! = \!\! & 4.1 \pm 7.0,\! 5.2 \pm 6.8  \label{eq:Kexcess} \eea
The second set of numbers is obtained using the smaller correlated $\Xi_c$ yield, which is probably more appropriate.  
A significant $3 \sigma $ $K^-$ excess remains.   The $K^-$ excess is also reflected in the total charged kaon excess,  which is a bit larger partly because the uncertainty in the total charged kaon multiplicity is smaller than the uncertainty in the individual $K^- $ multiplicity.   
Because of large uncertainties in the inclusive neutral kaon BR{'}s, the $K^0/\overline{K^0}$ result is consistent with either no kaon excess, or sizable kaon excess.  In the next section we discuss the impact of additional kaons from enhanced $b \to sg$.

\section{Kaons from enhanced $b \to sg $ }

We have studied kaon production from fragmentation in $\overline{B} \to X_{sg}$ decays using JETSET 7.4 with DELPHI tunings at order zero in $\alpha_s $.  The large energy release should make these decays well suited for a jet language description.   Initially, a string is stretched from the $s$ quark to the spectator quark via the gluon so that the gluon is a kink in the string carrying energy and momentum \ref{jetset}. 
Following the model of Ref. \ref{altarelli}, the $b $ quark is given a Fermi momentum $p_b$ which satisfies a Gaussian distribution,
\beq \Phi (\vec{p_b} ) = {4 \over {\sqrt{\pi} p_F^3 }} e^{- \vec{p_b}^2 
/p_f^2 }.\eeq
Fits to CLEO $B \to X \ell \nu $ data give $p_F = 270 \pm 40$ MeV \ref{CLEOAlam}.  
The $b$ quark mass is given by 
\beq m_b^2 = m_B^2 + m_q^2 - 2 m_B \sqrt{\vec{p_b}^2 + m_q^2}, \eeq
where $m_q$ is the spectator quark constituent mass, which we take to be $300$ MeV, and $m_B$ is the $B$ meson mass.  In the $b$ quark rest frame the $s$ and $g$ initially move back - to - back with energy $m_b /2$.  Fragmentation is carried out in the $B$ rest frame, and the resulting kaons are subsequently boosted to the $\Upsilon (4S)$ rest frame.  Although the 
$b$ quark kinetic energy has negligible effect on the total kaon yields 
it does reduce the kaon yields in the high momentum end point region.

The following kaon multiplicities are obtained per $\overline{B} \to X_{sg}$ decay, 
\beq K^- :K^+ :\overline{K^0} :K^0 :: 67\% : 19\% : 62\% : 15\%   .\eeq
As an illustrative example of the impact on the charged kaon excess we take ${\cal B}(\overline{B} \to X_{sg}) = 15\% $. 
Subtracting the corresponding kaon yields from Eq. 18
gives [\%]
\bea \Delta {\cal B}(\overline{B} \to K^- X)
 &=  & 5.8 \pm 5.8,\;6.9 \pm 5.8 \nonumber\\
\Delta{\cal B}(\overline{B} \to K^+ X) & =  & -1.8 \pm 4.2,\; -1.8 \pm 4.2 \nonumber\\
\Delta {\cal B}(\overline{B} \to K^+ /K^- X) & = & 4.0 \pm 5.4,\; 5.1 \pm 5.3 \nonumber \\
\Delta {\cal B}(\overline{B} \to K^0 /\overline{K^0} X)& = & -7.5  \pm 7.0,\; -6.4  \pm 6.8 .\label{eq:Kbsgexcess} \eea
According to this example the observed inclusive charged and neutral kaon yields can be accounted for at the $1 \sigma $ level if $b \to sg$
is substantially enhanced by new physics.

Next, we turn our attention to the kaon momentum spectrum.
In Fig. 1a we compare the $K_s$ momentum spectrum for ${\cal B}(\overline{B} \to X_{sg}) = 10\%$, obtained from the DELPHI tuned JETSET 7.4 Monte Carlo described above, with the measured spectrum from $B$ decays \ref{ARGUSKaons} in the $\Upsilon (4S)$ rest frame.  
Also included is 
the spectrum obtained from the SLD implementation of the CLEO $B $ decay Monte Carlo for $b \to c$ transitions.  
This simulation has been somewhat artificially tuned by the SLD collaboration to maximize the charged kaon yield and does 
not yet include $\overline{B} \to D \overline{D} X$ decays. From the figure it is clear that detecting or bounding kaon production from enhanced $b \to sg$ poses a severe experimental challenge.  Most of the kaons would be soft, possesing momenta typical of kaons from $b \to c $ transitions; the ratio of signal to standard model background for typical $K_s$ momenta would be about 1 part in 5 to 10.  

\begin{figure}[htb] \centerline{\psfig{figure=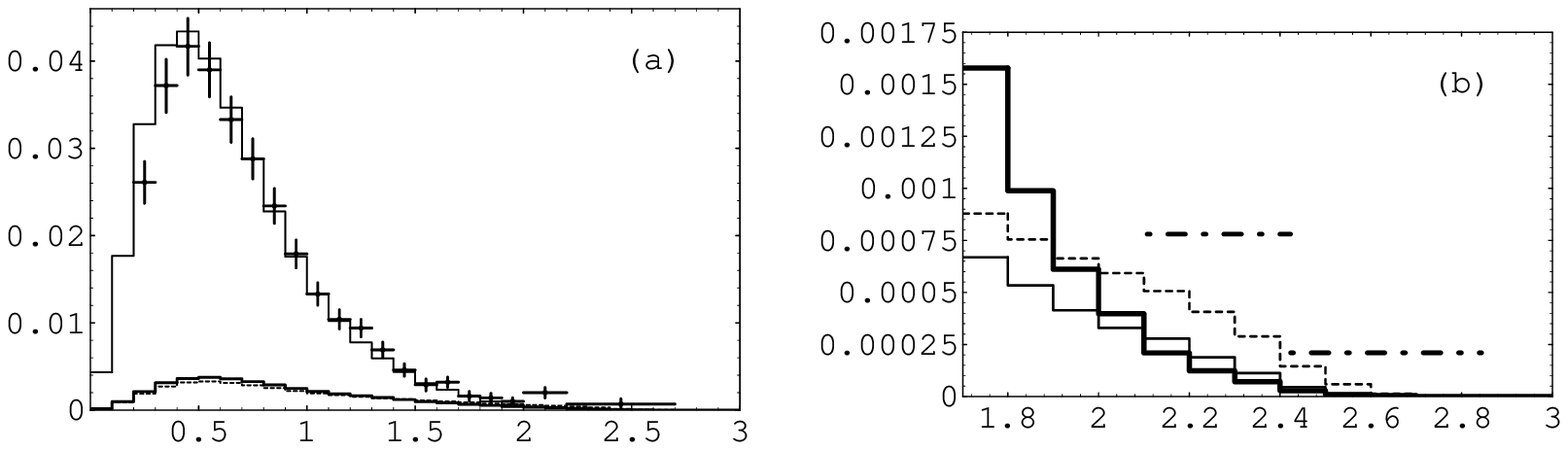,height=3.2in,width=7.00in}}{\label{fig:fig1} 
Figure 1: ${\cal B}(B \to K_s X)$ vs. $p_{K_s}$ [GeV]. Branching ratios are for $0.1$ GeV bins except CLEO upper limits.  
(a) ARGUS data (crosses), 
SLD Monte Carlo (top solid), Monte Carlo for ${\cal B}(\overline{B} \to X_{sg}) = 10\%$ with $p_F = 250$ MeV 
(bottom solid) and $p_F =0$ (dashed).  
(b) fast kaon spectra: CLEO 90\% CL UL{'}s for  $2.11 < p_{K_s} < 2.42$, $2.42 < p_{K_s} < 2.84$ (dot -dashed), 
SLD Monte Carlo (thick solid), Monte Carlo for ${\cal B}(\overline{B} \to X_{sg}) = 10\%$ with $p_F = 250$ MeV (solid)
 and $p_F =0$ (dashed).}\end{figure}

The signal to background ratio improves dramatically for high momenta where kaon production from $b \to c$ decays tends towards zero \ref{milana}.  
The CLEO collaboration has recently obtained upper bounds on $K_s$ production for $p_{K_s} > 2.1 \;GeV$ which are listed in Table 6 \ref{CLEOKsep}.  
The Monte Carlo yields for ${\cal B}(\overline{B} \to X_{sg}) = 10\%$ in Table 6 are consistent with the CLEO limits.  
Because there is currently no detailed data on kaon production from hadronic $Z$ decays for  $p_K /p_{beam} > .8$ with which to further tune JETSET, 
the actual high momentum kaon branching fractions per $b \to sg$ decay could certainly be 20\% smaller than DELPHI tuned JETSET 7.4 predictions.  
In addition CLEO may have problems with continuum over - subtraction at sensitivities of ${\cal O}(10^{-4} )$ for ${\cal B}(B \to K_s X)$.\footnote{For example,
the sum of the first two CLEO upper limits in Table 6 is 30\% larger than the third which could be a reflection of this.} 
It is therefore safe to say that ${\cal B}(\overline{B} \to X_{sg}) $ could be as large as 15\% based on the current $K_s $ bounds.
It will be interesting to see what CLEO's high momentum $K_s$ yields will be with the newly installed vertex detector in use for continuum subtraction.

\begin{table}\begin{center}
\caption{CLEO branching fraction upper limits [90\% c.l.] on $B \to K_s X$ ($\times 10^{4}$) at large kaon momenta, and corresponding lowest order Monte Carlo predictions for ${\cal B}(\overline{B} \to X_{sg}) = 10\%$ and $15\%$ with $p_F = 250\;MeV$.}
\vspace*{0.5cm} 
\begin{tabular} {| c | c | c | c |} \hline
$p_{K_s} [GeV] $  & CLEO UL  &  ${\cal B}(\overline{B} \to X_{sg}) = 10\%$  &  ${\cal B}(\overline{B} \to X_{sg}) = 15\%$ \\
\hline
2.11-2.42 & $7.8$ & $5.7$ & $8.5$  \\
2.42-2.84 & $2.1$ & $.5$ & $.7$  \\
2.11-2.84 & $7.5 $ & 6.2 & 9.2 \\
\hline
\end{tabular}
\end{center}
\end{table}

Unfortunately, the standard model background at large kaon momenta
due to $b \to s \bar q q$ penguin operator decays is subject to large theoretical uncertainty.  
Factorization model estimates\footnote{We would like to thank A. Dhatta for informing us of the results of Ref. \ref{Dhatta}.} 
\ref{Dhatta} of high momentum kaon production from penguin operator decays, where the kaons are formed from the primary
quarks in the decay,
give a direct $B \to K_s X$ branching ratio of about $10^{-4}$, with $K_s$ momenta above 2.1 GeV; $K_s$ production from $B \to K^* X$ decays is about three times larger, with a lower momentum distribution peaked
near 2.1 GeV.\footnote{Most kaons produced via penguin operator decays will be soft as in 
the case of $b \to sg$ decays due to energy degradation in the fragmentation process, leading to $B \to KX$ branching ratios at the 1\% level.}  
Since the factorization model estimates can not be trusted to better than a factor of two and since they are already of same order as the CLEO bounds it is not very useful to limit the search for enhanced $b \to sg$ to kaon momenta greater than 2.1 GeV.
Furthermore, if $b \to sg$ is enhanced, gluon splitting into light quark pairs at ${\cal O}(\alpha_s )$ leads to new contributions to kaon production which interfere with 
penguin operator contributions \ref{kaganperez}.
Interference in the factorization model can be substantial and destructive or constructive depending on the phase of 
the chromomagnetic dipole operator coefficient, introducing additional theoretical uncertainty at large kaon momenta.  
It is for this reason that we did not include parton showers in the $b \to sg$ Monte Carlo.   
Finally, there can be interference in various exclusive channels between high momentum kaon production from effective four 
quark operators (penguin operators, or gluon splitting in enhanced $b \to sg$ decays) and high momentum kaon production from fragmentation of 
enhanced $b \to sg$ at order zero in $\alpha_s$.

The above theoretical uncertainties all involve contributions to the $B \to K_s X$ branching ratio at the $10^{-4}$ level.  
Searches for enhanced $b \to sg$ at large kaon momenta would therefore have to
include a wider range of momenta corresponding to branching ratios at the $10^{-3}$ level.  For example, according to Fig. 1b, for $p_{K_s} \ge 1.8\;GeV$
as ${\cal B}(\overline{B} \to X_{sg})$ varies from 
10\% to 15\% the corresponding Monte Carlo contribution
to ${\cal B}(\overline{B} \to K_s X) $ varies from $1.9 \times 10^{-3} $ to $2.9 \times 10^{-3}$ for $p_F = 250$ MeV.  The Monte Carlo background from $b \to c$ decays is about $2.4 \times 10^{-3}$ so that the signal to background ratio would be roughly 1 to 1.\footnote{Although the Monte Carlo background does not yet include $\overline{B} \to D \overline{D} X$ decays, these will not substantially alter the signal to background ratio.}  Sufficiently precise knowledge of the dominant background contributions becomes a purely experimental issue rather than an intractable theoretical problem.
A critical analysis combining existing measurements of inclusive $D/D_s $ momentum spectra in $B $ decays \ref{CLEODspectra} , inclusive kaon momentum spectra in $D/D_s$ decays \ref{MARKIII}, and relevant $B$ decay Monte Carlo tunings can be used to determine how much uncertainty currently exists.  Of course, the uncertainty should be substantially reduced by future measurements at CLEO, BaBar, BELLE, and BES.
Finally, we note that although the above discussion considered $K_s$ production, essentially the same conclusions apply to charged kaon production.

\section{Discussion}

Updated measurements of the average $B$ decay charm multiplicity and ${\cal B}(\overline{B} \to X_{c \bar c s})$ at the $\Upsilon(4S)$
imply ${\cal B}(B \to X_{no\;charm} ) \approx 12.4  \pm 3.4  \pm 4.4 \%$.  
The uncertainty due to the $D$ and $D_s$ branching scales, i.e., $D^0 \to K^- \pi^+  $, $D^+ \to K^- \pi^+  \pi^- $, and $D_s \to \phi \pi $, 
has been separated out in the first error.  Hopefully, it will be significantly reduced by the BES collaboration or a future tau-charm factory.  
The second error should be substantially reduced at the $B$ factories.  
In the meantime, this result could be hinting at an ${\cal O}(10 \%) $ $b \to sg$ branching ratio due to the intervention of new physics, 
albeit at the $2 \sigma $ level.

Another hint for enhanced $b \to sg$ discussed in this paper comes from the $3 \sigma - 3.5 \sigma$ excesses in 
the inclusive $\overline{B} \to K^ - $ and $\overline{B} \to K^ + /K^- $ multiplicities beyond conventional sources: about $16.9 \pm 5.6 \% $ 
for $K^-$ and $ 18 \pm 5.3 \%$ for $K^+ /K^- $ .  Only taking the experimentally determined kaon yields from conventional sources into account 
leaves $4.4 \sigma $ and $5.6 \sigma $ excesses, respectively.  The largest unmeasured standard model contributions to kaon production, 
from $s \bar s$ popping in $\overline{B} \to X_{c \bar u d}$ decays, have been estimated using JETSET 7.4, both with default and DELPHI tunings.  
In principle these can be pinned down experimentally via
measurements of ${\cal B}(\overline{B} \to D K \overline{K} X)$, and improved measurements of ${\cal B}(\overline{B} \to D_s^+ K X)$.  
We have also added generous estimates for kaon yields from intermediate charmed baryon and charmonium decays.  

Two observations strongly suggest that the kaon excess is not due to unmeasured conventional sources.  First, 
the estimated central value of their contribution to the $K^- $ yield is about one third of the final $K^- $ excess.   
Second, about 90\% of the uncertainty in the $K^-$ and $K^+ /K^- $ excesses originates in the inclusive $B  \to K X$ measurements 
and in the experimentally determined $B \to D, D_s \to K $ yields.   Ultimately, these two sources of uncertainty will be substantially 
reduced at the $B$ factories, in combination with improved knowledge of the $D$ and $D_s $ branching scales.  
To date the ARGUS collaboration has presented the most precise measurements \ref{ARGUSKaons} 
of ${\cal B}(\overline{B} \to K^- X)$, ${\cal B}(\overline{B} \to K^+ X)$ and ${\cal B}(\overline{B} \to K^0 /\overline{K^0} X)$.  
It is therefore imperative that the CLEO collaboration update their 1986 measurements of these branching ratios \ref{CLEOKaons}, 
which are considerably less precise.

We have analyzed kaon production from fragmentation in $\overline{B} \to X_{sg}$ decays using JETSET 7.4 with DELPHI tuning.  
Our conclusion is that if ${\cal B}( \overline{B} \to X_{sg}) \sim 15 \% $ the additional kaon yield would reduce the $K^- $ and $K^+ / K^- $ 
excesses to $1 \sigma$ while maintaining $1 \sigma $ agreement with the total neutral kaon multiplicity.  
Furthermore, the associated $K_s $ momentum spectrum would be consistent with recent CLEO upper limits \ref{CLEOKsep} 
on $K_s$ production in the end point region, i.e., $p_{K_s} \ge 2.1 \;GeV$.  
Similarly, the corresponding $\phi$ momentum spectrum would be consistent \ref{kaganperez} with CLEO upper 
limits on $\phi $ production in the end point region \ref{CLEOphiep}.  

Although JETSET 7.4 with DELPHI tuning does quite a 
reasonable job of reproducing the observed kaon spectrum in hadronic $Z$ decays, more detailed data is needed at high kaon momenta, 
i.e., $p_K /p_{beam} > .8$.  
Obviously, for the purposes of studying $b \to sg$ it would also be useful to further tune JETSET using continuum kaon and $\phi$ 
production at the $\Upsilon(4S)$.  It is worth noting in this regard
that the BaBar and BELLE vertex detectors should be able to cleanly separate out kaons produced from the continuum charm component, 
leading to improved knowledge of kaon production from the strange component.

Kaons produced via ${\cal O} (10\%) $ $b \to sg$ branching ratios would account for about $10\%$ of the total kaon yield in $B$ decays.  Nevertheless, we have seen that their detection poses a formidable challenge because their momenta would populate the same region as kaons produced from intermediate charmed hadron decays, see Fig. 1.  In particular, the ratio of signal to standard model background would be about 1 part in 5 to 10 for typical momenta, e.g., $p_{K_s} < 1\;GeV$ in the $\Upsilon(4S)$ rest frame.  Excellent vertex detection will obviously be required in order to resolve the presence of charm decay vertices in background $B \to K$ decays with anything approaching the efficiency required at low kaon momenta.  The $B$ and $\overline{B}$ decay vertices need to be well separated, which of course will be the case at BaBar and BELLE.  It should be possible to further discriminate between signal and background by taking advantage of the back-to-back jet-like geometry of $b \to sg $ decays versus the more spherical geometry of $b \to c$ decays.

Finally, we have argued that searches for kaons in the end point momentum region where the $b \to c$ background tends to zero, e.g.,  $p_{K_s} > 2.1 \;GeV$ in the $\Upsilon (4S)$ rest frame, can not place very meaningful constraints on enhanced $b \to sg$ because of large theoretical uncertainty in the standard model background from $b \to s $ penguin operator decays, and the possibility of substantial destructive interference between the two contributions at large kaon momenta.  
Similar conclusions apply to $\phi$ production from enhanced $b \to sg$ \ref{kaganperez}.  These searches essentially probe $B \to KX$ branching ratios of order $10^{-4}$.  Instead, we propose that searches which attempt to take advantage of lower standard model backgrounds should include lower kaon momenta, e.g., $p_{K_s} \gtap 1.8 \;GeV$.  In this case one is probing $B \to KX$ branching ratios of order $10^{-3}$.  For ${\cal B}(\overline{B} \to X_{sg}) \sim 10\%$ to $15\%$ the signal to background ratio would be about 1 to 1, as in searches restricted to larger kaon momenta, but the standard model background would be dominated by $b \to c$ decays.  The advantage is that 
the latter can be pinned down experimentally, unlike the background from
penguin operator decays.  A dedicated effort should be undertaken 
in this regard to measure the inclusive $B \to D/D_s $ and $D/D_s \to K$ momentum spectra as accurately as possible.  
In addition, vertex detectors would not have to be nearly as efficient as at lower more typical kaon momenta in order to achieve 
useful reductions in the $b \to c$ background. 

\bigskip

\vskip.25in
\centerline{ACKNOWLEDGEMENTS}

It is a pleasure to thank Tom Browder, Alakabha Datta, Su Dong, Isi Dunietz, Adam Falk, Susan Gardner, Klaus Hamacher, Klaus Honscheid, Randy Johnson, Mike Luke, Antonio Perez, Mike Sokoloff, Mikhail Voloshin, and Hitoshi Yamamoto for useful conversations.

\bigskip

\newpage
\renewcommand{\baselinestretch}{1}

\end{document}